%% file: ms.tex
\documentclass[journal]{vgtc}         

\ifpdf
  \pdfoutput=1\relax                   
  \usepackage{graphicx}                
  \DeclareGraphicsExtensions{.pdf,.png,.jpg,.jpeg} 
\else
  \usepackage{graphicx}                
  \DeclareGraphicsExtensions{.eps}     
\fi%

\graphicspath{{figures/}{pictures/}{images/}{./}} 

\usepackage{microtype}                 
\PassOptionsToPackage{warn}{textcomp}  
\usepackage{textcomp}                  
\usepackage{mathptmx}                  
\usepackage{times}                     
\usepackage{cite}                      
\usepackage{tabu}                      
\usepackage{booktabs}                  
\usepackage{amsmath}    
\usepackage[disable]{todonotes}
\usepackage{algorithm}
\usepackage{algorithmic}
\usepackage{balance}

\newcommand{\ds}[1]{{#1}}
\newcommand{\ac}[1]{\todo[inline,color=blue!20!white]{\textbf{1AC:} #1}}

\onlineid{1730}

\vgtccategory{Research}
\vgtcpapertype{Evaluation}

\title{The Weighted Average Illusion:\\Biases in Perceived Mean Position in Scatterplots}

\author{Matt-Heun Hong, Jessica K. Witt, and Danielle Albers Szafir \textit{Member, IEEE}}
\authorfooter{
\item
 Matt-Heun Hong and Danielle Albers Szafir are with the ATLAS Institute, University of Colorado Boulder. Email: matt.hong@colorado.edu, danielle.szafir@colorado.edu.
\item
 Jessica K. Witt is with the Department of Psychology, Colorado State University. E-mail: jessica.witt@coloradostate.edu.
}

\shortauthortitle{Hong \MakeLowercase{\textit{et al.}}: The Weighted Average Illusion}

\abstract{Scatterplots can encode a third dimension by using additional channels like size or color (e.g. bubble charts). We explore a potential misinterpretation of trivariate scatterplots, which we call the \emph{weighted average illusion}, where locations of larger and darker points are given more weight toward x- and y-mean estimates. This systematic bias is sensitive to a designer’s choice of size or lightness ranges mapped onto the data. In this paper, we quantify this bias against varying size/lightness ranges and data correlations. We discuss possible explanations for its cause by measuring attention given to individual data points using a vision science technique called the centroid method. Our work illustrates how ensemble processing mechanisms and mental shortcuts can significantly distort visual summaries of data, and can lead to \ds{misjudgments} like the demonstrated weighted average illusion.
} 


\CCScatlist{ 
 \CCScat{K.6.1}{Management of Computing and Information Systems}%
{Project and People Management}{Life Cycle};
 \CCScat{K.7.m}{The Computing Profession}{Miscellaneous}{Ethics}
}

\teaser{
  \centering
  \includegraphics[width=\linewidth]{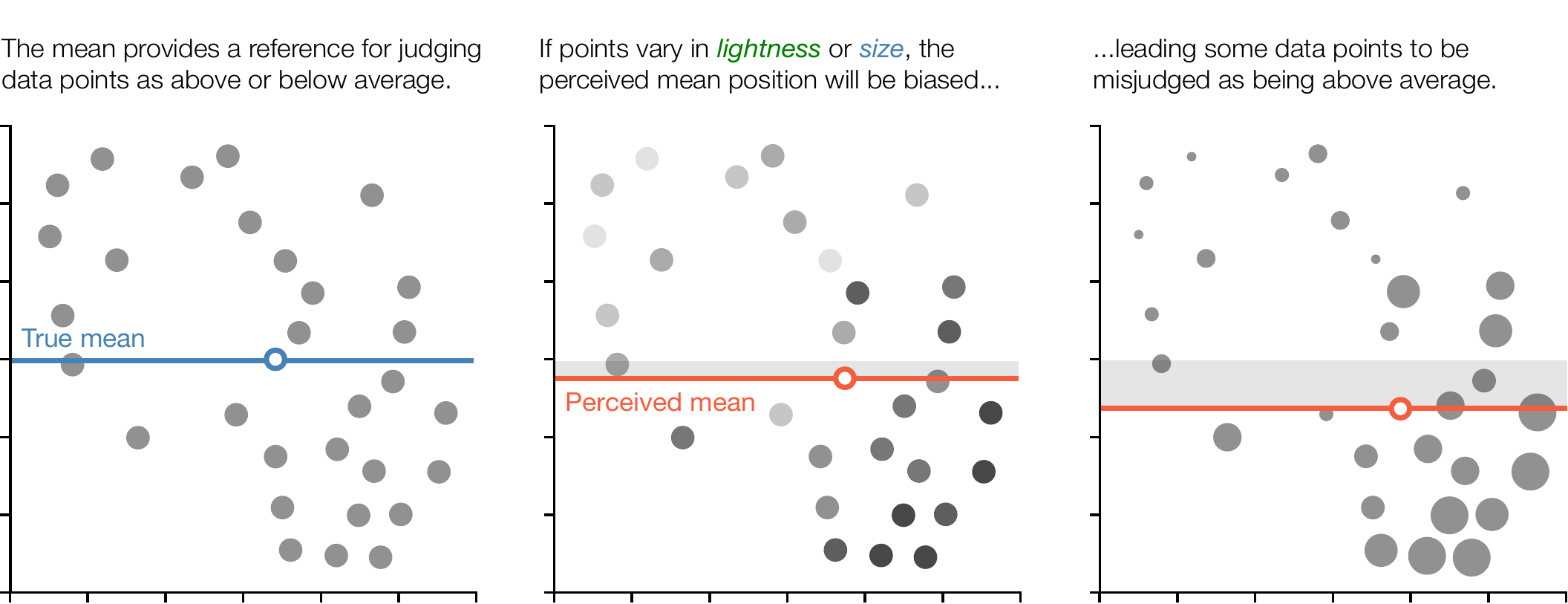}
  \caption{We measured how integrating size or lightness into scatterplots can systematically bias the perceived mean of those points. We model how individual marks contribute to this bias, allowing us to predict differences in the actual mean (blue) and where people see the mean (orange). This misinterpretation can inhibit sensemaking and decision making. For instance, points in the grey boxes above will be incorrectly perceived as having higher-than average y-values. 
  	Under this illusion, graph readers' mean estimates can be increasingly biased as a function of size or lightness ranges mapped onto the data. }
  \label{fig:teaser}
}

\vgtcinsertpkg

\begin{document}

\maketitle
\begin{figure}[tb]
	\centering
	\includegraphics[width=\columnwidth]{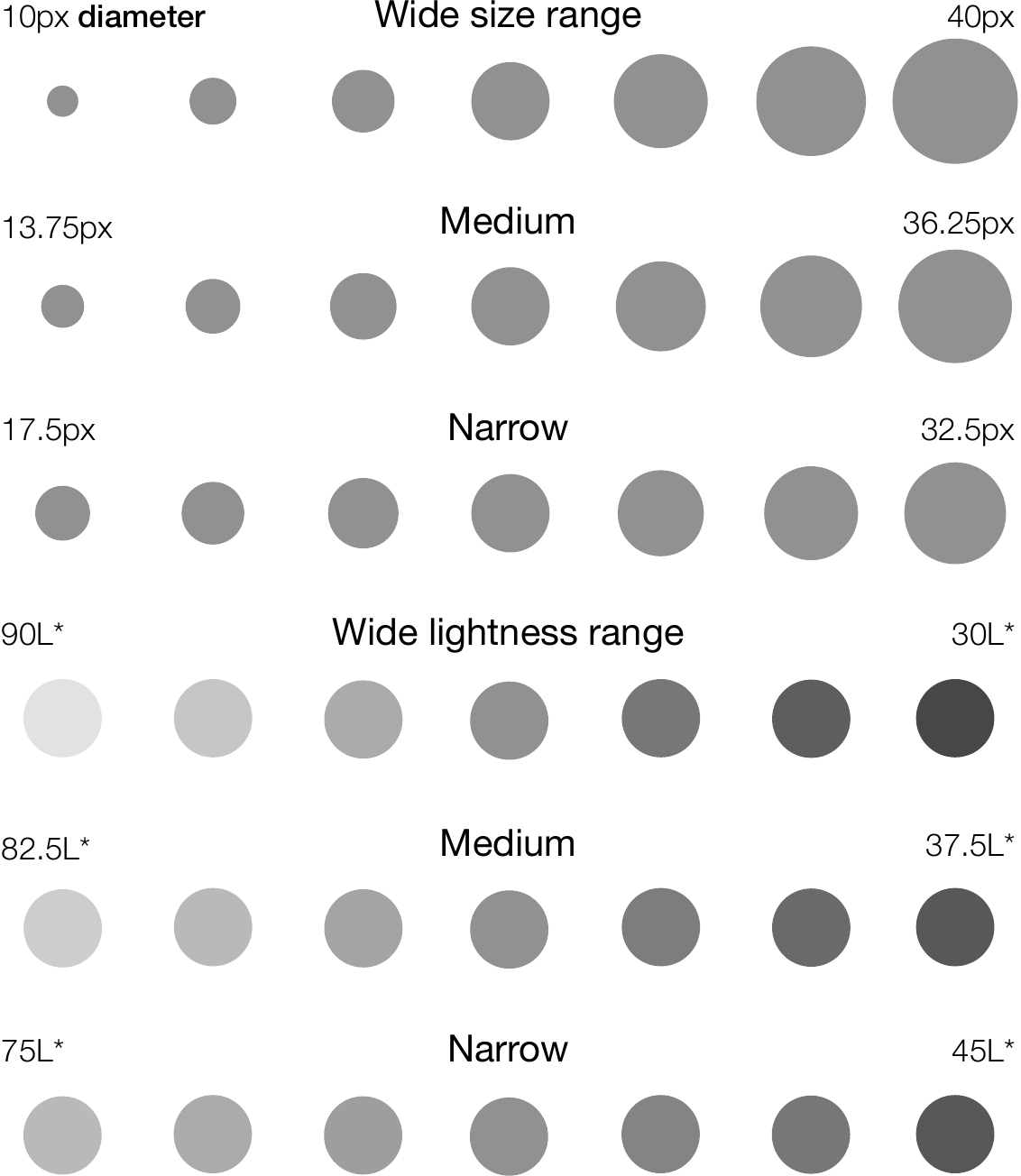}
	\caption{We mapped the third dimension in trivariate scatterplots to one of three size or lightness ranges (shown here to scale). Steps in pixel diameter and L* values are evenly spaced, with the midpoints of all ranges being identical. Every scatterplot stimulus contained seven different mark sizes or lightness.}
	\label{fig:ranges}
\end{figure}

\section{Introduction}
\input{intro}

\section{Background}
\input{background}

\section{Hypotheses}\todo{Typo [R2, R3]}
\input{hypothesis}

\begin{figure*}[h!]
	\centering
	\includegraphics[width=\textwidth]{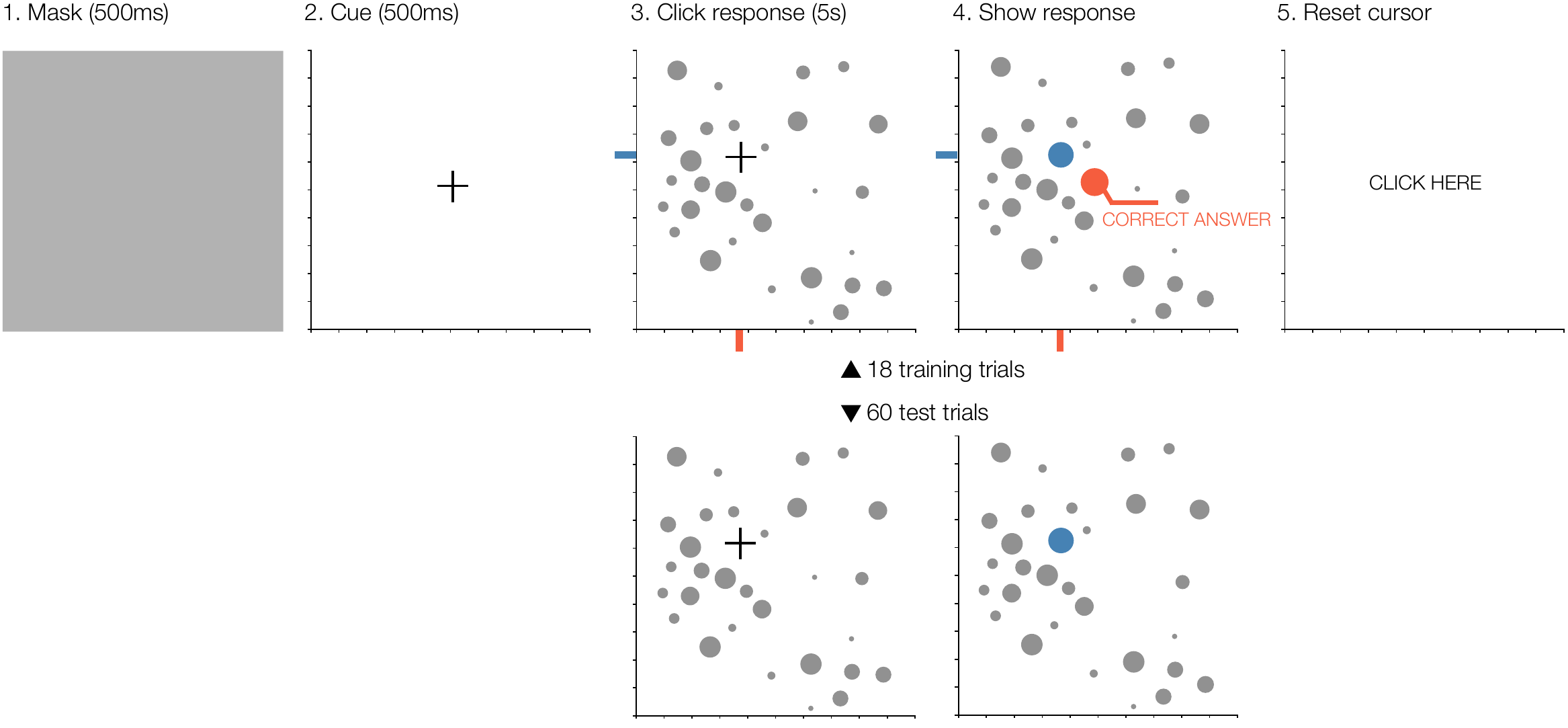}
	\caption{The formal study consisted of five steps. A grey mask (1) would be replaced by a fixation cross (2) to guide participant attention. Participants would then see a scatterplot (3) and report the perceived mean position by clicking on the corresponding location in the scatterplot, with the cross moving with the cursor. We would visually indicate the reported value (4) and then ask participants to click on a link at the center of the screen (5) to recenter their mouse before the next trial.}
	\label{fig:procedure}
\end{figure*}  

\section{Methods}
\input{methods}

\section{Results}
\input{results}

\section{Modeling Feature-Based Attention}
\input{attention}

\section{Discussion}
\input{discussion}

\subsection{Limitations \& Future Work}
\label{sec:limitations}
\input{limitations}
\section{Conclusion}
\input{conclusion}

\acknowledgments{
This work was supported in part by NSF awards BCS-1632222, SES-2030059, IIS-2046725, IIS-1764092, and IIS-1764089.
}

\balance
\bibliographystyle{abbrv-doi}
\ac{Clarified citation [6]}
\bibliography{bib}
\end{document}

%% file: intro.tex

\ac{[R2] We have restructured the first four paragraphs to be a more intuitive reading for visualization researchers. This includes changes in vocabulary, e.g. changing 'features' to 'channels', and changing 'objects' to 'point marks'.}

Effective visualizations like scatterplots communicate data by leveraging fast and accurate visual processes.
Scatterplots map two data dimensions to position \cite{Sarikaya, Bertini2021}, a precise channel for comparing values \cite{Cleveland1984,albers2014task,Kim2018c}. 
They also leverage our ability to summarize sets of point marks through \emph{ensemble processing} mechanisms \cite{ariely2001seeing,haberman2012ensemble,Szafir2016d,whitney2018ensemble}. 
Ensemble processing helps readers easily intuit how data points are distributed,
allowing judgments about summary statistics such as correlations \cite{harrison2014ranking,Rensink2010a}, position means \cite{Gleicher2013d,Wei2019}, and clusters \cite{abbas2019clustme}. 

While ensemble perception is fundamental to visual data comprehension, it has limitations that can also interfere with effective communication. Recent work in vision science suggests that visual channels corresponding to common design elements, like size or color, may systematically bias our abilities to estimate the mean position of a small collection of point marks \cite{Drew2010c,Sun2016c,Rodriguez-Cintron2019c}. These limitations could manifest in common visualization techniques like line charts---which lead to underestimation of means \cite{Xiong2020c} and overestimation of trends \cite{WittWarden2020})---and scatterplots.

Our work focuses on a systematic bias in estimating the mean position of data points in a scatterplot. This task is commonly used to assess differences across groups of data points \cite{Sarikaya, Gleicher2013d}. For example, Rosling used multiclass scatterplots for Gapminder \cite{rosling2011health} to compare mean GDP per capita and mean mortality rates between geographic regions. Using small multiples scatterplots like Parlapiano \cite{parlapiano_2014}, people can assess trends in per capita income and life expectancy by comparing means and variances across time \cite{kramer2017visual}. Mean position can also provide a reference point: a survey \cite{hessney_peck_dickensheets_2020} published by The New York Times used a scatterplot to ask readers which venues should reopen first during the COVID-19 pandemic. Readers' decision criteria \cite{Stewart2000} are likely to be determined based on the perceived x- and y-means, as illustrated in Hessney et al. \cite{hessney_peck_fetter_2019} and in Figure \ref{fig:teaser}.

%

We conducted a crowdsourced study with 130 participants (\S\ref{sec:stim}) to measure biases in mean position estimates in scatterplots where point marks varied in size or lightness. Our study illustrates graph readers' enduring tendency to weigh large or dark marks more toward the mean. However, this is a misinterpretation of scatterplots, where weighted position means will shift with the range of size or lightness mapped onto the data (see Figures \ref{fig:teaser} and \ref{fig:ranges}).

Our demonstration of this \emph{weighted average illusion} and the ensuing bias adds to the growing literature on visualization biases \cite{dimara2018task, Xiong2020c, WittWarden2020}. To guide our discussion of causal factors in \S\ref{sec:strategies}, we incorporate a statistical model of feature-based attention (FBA) \cite{Carrasco2011} known as the \emph{centroid method} \cite{Sun2016c} which quantifies attention given to individual data marks (\S\ref{sec:centroid}). In addition, our models lead to hypotheses about other strategies people might utilize, such as selective attention \cite{Myczek2008} (\S\ref{sec:subsampling}) and spatial segmentation \cite{Franconeri2009c}, that may help practitioners understand how similar biases may arise in other designs (\S \ref{sec:density}).

\vspace{3pt}
\noindent \textbf{Contributions: }
Our primary contribution is quantifying the bias that may arise under the weighted average illusion in trivariate scatterplots, as a function of design choices and data patterns (Figure \ref{fig:pulls}), even after training. We also demonstrate how modeling techniques in visual cognition like the centroid method can enable causal interpretations of visualization studies. These models allow practitioners to predict when biases may arise and avoid potential misjudgments. Our findings create opportunities at the intersection of visualization and vision science by \ds{hypothesizing possible} mental shortcuts used in visualization interpretation and decision making.

%% file: background.tex
Honest communication in visualization means avoiding deceptions and biased interpretations of data. Understanding bias has a long tradition in visualization research \cite{pandey2015deceptive,correll2017black,szafir2018good,mcnutt2020surfacing, witt2019graph}, and one recent study demonstrates possible priming effects on position means \cite{Xiong2020c}. We suspect that position mean biases could also be caused by limits on our visual system. We build on knowledge and techniques from vision science to model bias as a function of visual elements in scatterplots.

\subsection{Visualization Biases}
Visualizations provide a powerful yet imperfect means for communicating data. Subtle design choices may bias conclusions drawn from data \cite{correll2017black,mcnutt2020surfacing}. For example, encoding data using a bubble's area rather than its diameter can inflate perceived data differences \cite{pandey2015deceptive}. Other biases may emerge from choices made in the data processing pipeline. For example, changing the rendering order of scatterplot points can distort our perception of distributions \cite{mcnutt2020surfacing}. And cognitive biases often do not stem from the visualizations themselves, but from imperfections in an analyst's sensemaking process (see Dimara et al. \cite{dimara2018task} for a survey).

Design guidelines can provide proactive strategies for avoiding common biases \cite{szafir2018good}. For example, designers are encouraged to avoid rainbow colormaps, partly because they can cause ``banding'' biases that lead people to group marks that share similar hues \cite{borland2007rainbow,quinan2019examining}. Intelligent design systems, such as visualization linters \cite{hopkins2020visualint,mcnutt2018linting}, can leverage these guidelines to identify potentially misleading practices.  

However, we lack formal models of biases that can help us reason about potential design trade-offs. While controlled experiments \cite{pandey2015deceptive} substantiate the harmfulness of distorting aspect ratios and truncating y-axes, this latter (much maligned) practice can be helpful for interpretation depending on the data patterns and communication goals \cite{correll2020truncating,witt2019graph}. If magnitudes of bias could be modeled as a function of visual elements, designers can make better judgments about these trade-offs to enhance the overall effectiveness of a visualization while limiting bias \cite{ritchie2019lie}. 

While bias can emerge as an artifact of explicit design choices, other biases may be more subtle. Xiong et al. \cite{Xiong2020c} showed that position means can be biased in conventional plots: people systematically overestimated the mean using bar charts, and underestimated it using line charts. Statistical patterns in scatterplots can be distorted by geometric scaling \cite{Wei2019}. Visualizations optimized for one task may fail for many others \cite{Pinker1990}, and such failures may lead to bias: even the most honest charts might mislead the reader. 

\subsection{Scatterplot Perception}
Scatterplots are amongst the most commonly used \cite{russell2016simple} and scrutinized \cite{Sarikaya} visualization techniques. They primarily map data to both x- and y-positions, yet encompass a diverse range of design variations\todo{Clarification [R2]} \cite{Sarikaya} and are applicable to a broad array of tasks, from comparing individual values \cite{correll2012comparing,albers2014task,Cleveland1984} to summary statistical tasks such as clustering \cite{sedlmair2012taxonomy}, averaging \cite{Gleicher2013d}, and correlation \cite{harrison2014ranking,Rensink2010a}. 

Their variations \cite{Sarikaya}\todo{Clarification [R2]} frequently communicate more than two data dimensions. The ways these additional dimensions are depicted can impact the scatterplot's usability, making it easier or harder to estimate trends \cite{correll2017regression} or compute differences across groups of data points \cite{Gleicher2013d,burlinson2017open}. Computational models of these trade-offs exist for optimizing scatterplot designs for different tasks \cite{micallef2017towards}. 

Such trade-offs in scatterplot designs become more complex as a function of visual channels used to encode data. The term \emph{separability} refers to the ease with which our visual system can process one visual channel without interference from another \cite{Munzner2018c,ware2019information,garner1976interaction}. For example, color and position channels are separable, \ds{since our perception of position is robust to changes in color, and vice versa}. But when visualizing data using red hue values for one measure and blue hue values for another, people will struggle to process each measure independently \cite{ware2019information}. 

While position is generally considered separable from all other channels, evidence suggests that position may be integral with some channels for more complex visualization tasks \cite{correll2017regression,Kim2018c}. For example, position is integral with motion in outlier detection in multivariate scatterplots \cite{veras2019saliency}. Recent studies provide formal models of separability across color, shape, and size for comparing data values \cite{Smart2019b,demiralp2014learning,szafir2017modeling}, but we lack \todo{Typo [R3]}formal models for the separability of position with other channels in perceiving means and distributions. 

A scatterplot's ability to support averaging and other ensemble tasks may also be affected by the underlying data distributions \cite{Kim2018c, WittWarden2020}. While one study showed that averaging in scatterplots is generally robust to the size of data and a variety of tertiary encodings \cite{Gleicher2013d}, positional outliers may bias trend perceptions \cite{correll2017regression}, and performance on position summary tasks can vary across geometric scales \cite{Wei2019}. In this study, we model the separability of position and two common visual channels---size and lightness---for the position mean task across varying data distributions.

\subsection{Ensemble Perception}
Visualization tasks often involve extracting statistical summaries from data, such as mean position and size of data points; detecting color, size, or position outliers; spatial or feature-based segmentation; and judging correlation \cite{Szafir2016d}. 
These tasks rely on our ability to rapidly summarize sets of visual elements through \emph{ensemble processing} \cite{ariely2001seeing,haberman2012ensemble,Szafir2016d,whitney2018ensemble} at a glance, even without\ac{Clarification [R1]} focused attention to locations of individual objects \cite{Feigenson2011a, dakin1997computation,witt2019perceptual,Rensink2010a,hochstein2018comparing,Cohen2016a}. While ensemble perception is mostly beneficial for data comprehension, it may be susceptible to bias. Although mark color should be separable from position, Sun et al. \cite{sun2018high} suggests color lightness will systematically bias position averaging. These biases can lead to illusory misjudgments when viewing dot plots. 

Previous work \cite{Xiong2020c} studied\todo{Typo [R2]} the impact of higher-level cognitive influences on position mean biases. Our work draws on vision science techniques to model bias in terms of visual elements using the regression models of the \emph{centroid method} \cite{Sun2016c}. These models have been used to explain systematically biased mean estimates as a function of object lightness \cite{Sun2016c,sun2018high}, hue \cite{sun2016human}, size \cite{Rodriguez-Cintron2019c}, orientation \cite{inverso2016evidence}, and texture \cite{Sun2016c} in small sets of objects. However, these studies aim to isolate specific mechanisms, so stimuli are presented in subsecond durations. Only twelve or fewer targets are displayed, and no additional context such as labels or axes could divert attention. We extend their methods to realistic visualization scenarios to understand how biased data interpretations may arise as a function of design choices.

%% file: hypothesis.tex
Graph readers can use scatterplot means to characterize a range of values, compare classes in multiclass scatterplots \cite{Gleicher2013d}, compare distributions over time or other facets \cite{bertin2010graphische,kramer2017visual}, or create decision thresholds \cite{hessney_peck_fetter_2019, Stewart2000}. 
\todo{Removed a sentence}
Studies in vision science showed that common visual variables like lightness or size can interfere with people's abilities to estimate mean properties for a small collection of objects \cite{Sun2016c,Drew2010c,Rodriguez-Cintron2019c}. 
However, these studies sought to measure perceptual mechanisms using flash tasks (i.e., stimuli were displayed for 500ms) and asking people to average fewer than 16 randomly distributed objects. 
These interference effects may not hold in visualization contexts, \ds{where people see a large number of marks, including the axes and ticks, for longer periods of time.}
To measure potential systematic bias in scatterplot averaging, we asked participants to indicate the average position of all marks in scatterplots where a third data dimension was mapped to either lightness or size. 
Based on prior studies, we hypothesize:

\vspace{3pt}
\noindent \textbf{\emph{H1:} Scatterplot means will always be pulled towards locations of dark or larger points. }\\
Vision science studies demonstrate that locations of \ds{more salient} (larger or darker)\todo{Clarification [R1]} marks pull the perceived mean among small sets of objects. We expected these results to extend when reading trivariate scatterplots over the course of several seconds.\todo{Clarification [R2]}

\vspace{3pt}
\noindent \textbf{\emph{H2:} Increasing correlations between the irrelevant channel and position will increase bias. }\\
Increased correlation between position and either size or lightness will cause dark or large data points to group together. If locations of dark or large data points pull mean position, we will see significantly biased mean responses directed towards regions where these points are located.
\todo{Stylistic [R2]}Kim \& Heer\cite{Kim2018c} demonstrated increased error rates with scatterplots with size variance, suggesting that size may have a stronger biasing effect than lightness.  

\begin{figure}[!t]
	\centering
	\includegraphics[width=\columnwidth]{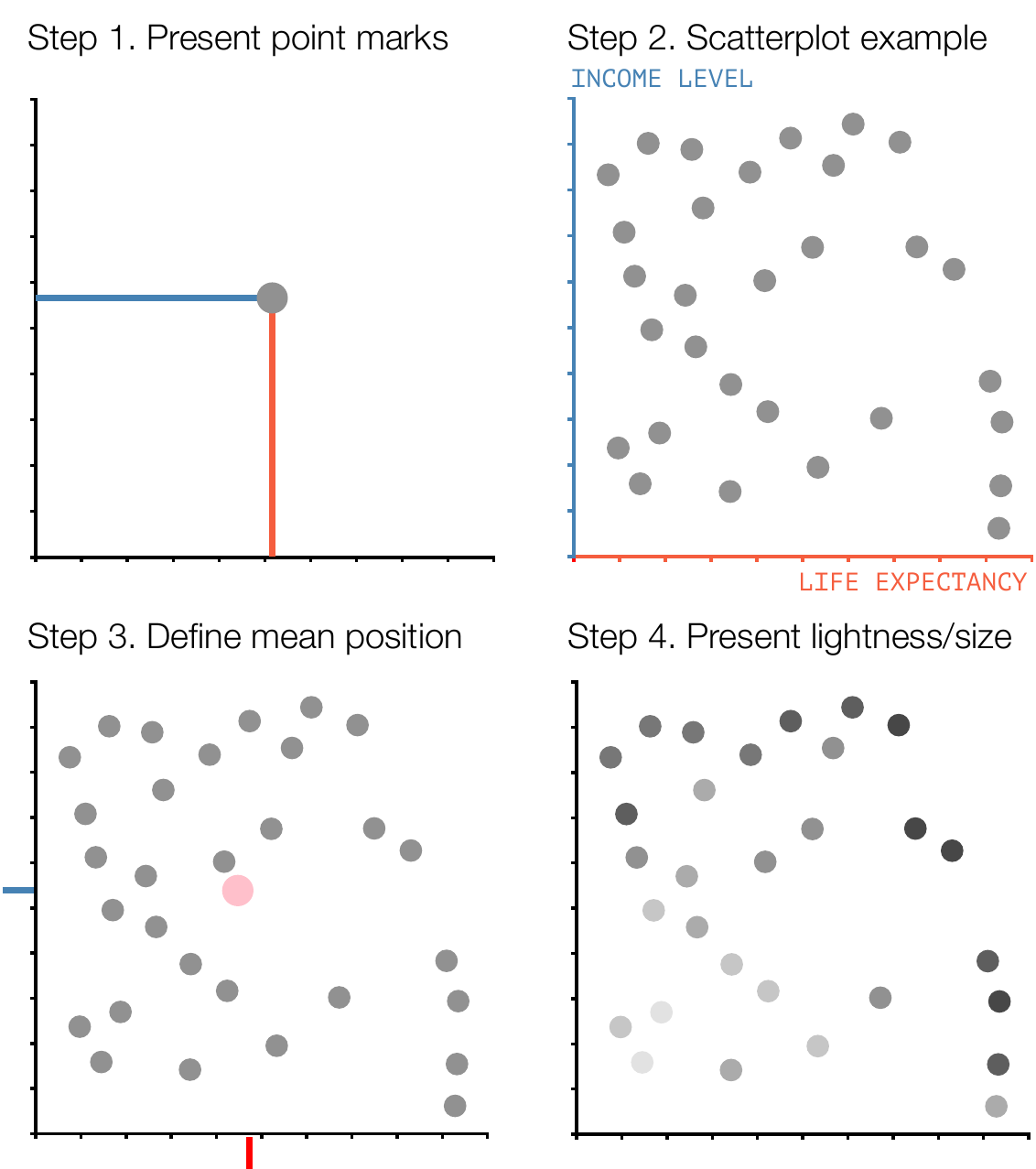}
	\caption{To ensure comprehension, we stepped participants through the target task, steadily building in more information. The tutorial images above were presented serially with the following instructions: 
		1. Scatterplots reduce information about two measures into a single data point. 2. On this scatterplot, 30 countries are presented in terms of their life expectancies and income levels. 3. Therefore, the pink dot alone represents both the average life expectancy and the average income of the 30 countries shown. 4. Each point on a scatterplot can also depict a third variable, such as unemployment rate. After viewing the tutorial, participants were instructed that: ``In the following study, you are asked to estimate and click on the average position of all points (i.e., average life expectancy and average income of all countries) on each scatterplot.''}
	\label{fig:tutorial}
\end{figure}

\vspace{3pt}
\noindent \textbf{\emph{H3:} Widening the encoding range of the irrelevant channel will amplify bias. }\\
Increasing the visual difference among data points increases the differences in contrast across a scatterplot (e.g., the large marks would become larger, and the small marks smaller). 
If variations in contrast already shift the perceived position mean, increasing this difference would increase the ensuing bias. 


%% file: methods.tex
We investigated the effects of size and color on mean position estimation for trivariate scatterplots in two separate experiments: one investigating size and the other lightness. 
Each experiment was a 3 (encoding ranges) $\times$ 3 (correlations) within-subjects design. 
We measured performance using the vector 
between the reported means and the true means. 
\ds{We provide anonymized data and our study infrastructure at \url{https://osf.io/h8ft3/?view_only=9564278544b4411c82610c73daed8c00}.}

\subsection{Stimuli Generation}
\label{sec:stim}

\ds{Our stimuli consisted of 500 $\times$ 500 pixel scatterplots generated
using D3 (Figure \ref{fig:tutorial}). 
Each scatterplot was rendered on two orthogonal black axes with unlabeled tick marks every 50 pixels.}


\ds{To generate the x- and y-data, we used Poisson disk sampling\cite{PoissonDisk} to produce 30 uniquely distributed point grids, with minimum distance between the boundaries of any two points set at 8 pixels. 
This methodology is similar to Gleicher et al. \cite{Gleicher2013d}.
Each dataset always contained 30 marks, with the number of points selected in piloting.}

\ds{For each of the above datasets, we generated additional data for size and lightness to satisfy each of three spatial correlation levels: no correlation, low correlation, and high correlation. 
In the no correlation condition, the position of points had no correlation with the size or lightness data, providing a random distribution of the distractor encodings. 
In the low correlation condition, both the x- and y-positions of points were correlated with size or lightness by $\rho = 0.4\pm0.05$; 
in the high correlation condition, position was correlated with size or lightness by
$\rho = 0.8\pm0.05$. 
We generated four datasets for each direction of correlation---either increasing or decreasing, along either the positive or negative diagonal---for each of the 30 point grids.
These correlations between the third measure and the two position measures created a visual gradient (from light-to-dark, or small-to-large) along one of these four diagonals (Figure 1, right).
This process led to 12 variations for each of the 30 scatterplot stimuli.}\ac{Clarification [R4]}

\ds{The encoding range for each distractor variable (size or lightness) corresponded to one of three levels: $45L^*$ to $75L^*$, $37.5L^*$ to $82.5L^*$, and $30L^*$ to $90L^*$ for lightness and 17.5px to 32.5px, 13.75px to 36.25px, and 10px to 40px for size, specified in terms of mark diameter\todo{Clarifying that we used diameter [R1]} 
	(Figure \ref{fig:ranges}). 
The linear\todo{Clarifying that these are linear scales [R1]} 
ranges were sampled at seven 
evenly spaced values and 
shared the same middle value (i.e., the same size or lightness) to focus on the effect of range width rather than its midpoint. 
We chose the lightness ranges based on conventional $L^*$ distributions in ColorBrewer \cite{ColorBrewer}, where the darkest colors are typically near 30L*, and the brightest colors are typically near 90L*. 
The ranges of size were restricted to those that would prevent occlusion given the 8 pixel limit on distance between points and align with default ranges found in commercial systems.
In both size and lightness\todo{Use 'lightness' instead of 'color' [R1]} 
experiments, the narrowest range spanned half the width of the largest range.}

\ds{Each participant saw 60 trials: 54 test trials varying in size or lightness and 6 control trials with no size or lightness variation to provide a baseline error rate. The test trials consisted of six trials for each combination of correlation (none, low, high) and encoding range (narrow, medium, wide). Data for each trial was randomly sampled from the set of pre-generated datasets with the correct corresponding correlation level. We presented the 60 trials in a random serial order.}
%
\ac{Paragraph clarification [R2, R3, R4]} 

\begin{figure*}[!ht]
	\centering
	\includegraphics[width=\textwidth]{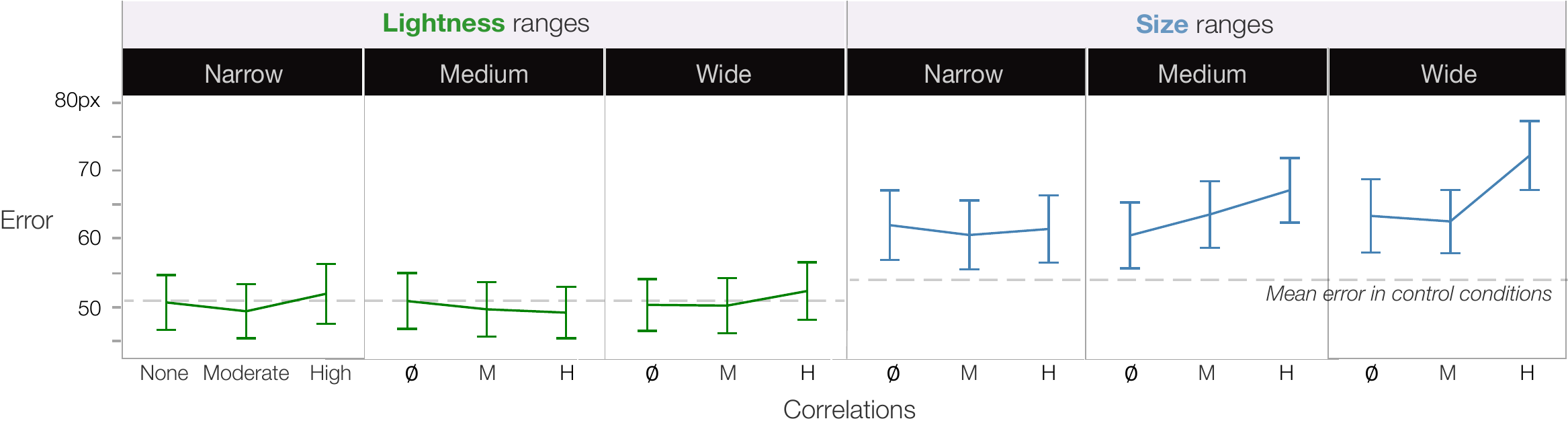}
	\caption{Mean error magnitude across encoding type, encoding range, and correlations. Note the y-axis represents pixel values, and our stimuli size was 500\textit{px} by 500\textit{px}. Using lightness (green lines) did not significantly increase error over the baseline. In the size experiment (blue lines), errors were significantly higher than the baseline condition for all ranges. Error bars represent 95\% confidence intervals.}
	\label{fig:errors}
\end{figure*} 
\ac{Noting the size of our stimuli [R4]}

\subsection{Procedure}
\label{sec:procedure}
We conducted a Mechanical Turk\todo{Clarification [R2]} 
experiment consisting of four phases: 1. informed consent, 2. instructions and tutorial, 3. formal trials, and 4. demographics. 
Participants were first provided with a consent form containing basic information about the data to be collected in the study. 
After providing consent, each participant passed an online\todo{Clarification [R2]} 
Ishihara plate test to screen for color vision deficiencies \cite{clark1924ishihara}.

After completing the screening, participants moved to a tutorial to ensure proper task understanding. The tutorial walked participants through the \ds{design of a trivariate scatterplot}, first describing mark position, then demonstrating mean mark position, and finally \ds{introducing an additional channel using unemployment rate as an example measure. \todo{Use 'lightness' instead of 'color' [R1]}  
The tutorial, their accompanying figures, and text annotations were iterated through extensive piloting. 
During in-person piloting, we debriefed participants on our research goals, and confirmed that no participant interpreted the additional third measure as a factor that should affect the mean value of scatterplots.
Figure \ref{fig:tutorial}\todo{Typo [R3, R4]} 
summarizes this aspect of the experiment.}
 

\ds{After the tutorial, participants were instructed to ``\textit{Click on the average position of all points}" on each scatterplot, and this instruction persisted throughout both training and test trials. Training consisted of 18 trials where participants received immediate feedback by being presented the true mean position (Figure \ref{fig:procedure}) after their click response. 
Prior work \cite{Sun2016c} utilized such feedback to collect more consistent estimates of mean position.
During these practice trials, cursor movements were animated by moving reference lines along the x- and y-axis, reinforcing the idea that the participant was to average the x- and y-positions of the data points.} 

During the test trials, the interactive cursor guides and feedback were removed.
Participants saw each of their 60\ac{Clarification [R2, R3, R4]}
scatterplots in a random serial order. 
Each trial began with a 500ms gray mask followed by a fixation cross symbol to cue the participant that the trial was about to begin. 
After another 500ms, the scatterplot was rendered and participants had five seconds to click on the mean position of points in the scatterplot, with the time limit determined in piloting. 
A pink dot appeared briefly at the clicked location, signaling the trial's end. 
Participants could complete the task after the scatterplot was hidden, but delayed responses prompted an alert to encourage the participant to respond within the allotted time.
Before beginning the next trial, participants had to move the cursor back to the center of the stimulus by clicking a link in the middle of the scatterplot to reset their cursor position, minimizing potential motor bias.

\ds{We interspersed four engagement checks in the formal study to assess honest participation. 
On average, 7.5 trials passed before the first engagement check, and then 15 trials passed before each successive check.\ac{Revision [R1, R4]}
During these engagement checks, a single data point was shown in one of the four quadrants, and participants were removed from the analysis if they failed two or more checks. 
After completing all trials, participants completed a demographic survey and were compensated for their participation.}

\subsection{Measures \& Analysis}
\label{measures}
Error was measured as the magnitude of the error vector between the true mean and the reported mean. We additionally computed the directional bias in responses by projecting the error vectors to the direction of the correlation gradient.

We analyzed these measures using a two-factor (encoding range, and correlation level) ANCOVA with trial order, direction of the correlation gradient, the specific datasets, and interparticipant variation as random covariates. 
We examined both primary effects and first-order interaction effects and used Tukey's Honest Significant Difference (HSD) test ($\alpha = .05$) with Bonferroni correction for post-hoc comparisons. 
These analyses excluded the control scatterplots (those with no lightness or size differences) as those trials reflect no measurable correlations or encoding ranges. 
However, we used Dunnett's Method to identify where error or bias significantly differed from baseline performance measured in these trials. 

\subsection{Participants}
We recruited 174 subjects with US IP addresses from Amazon's Mechanical Turk. We excluded 22 participants with limited or no cursor movements, flagged by back-to-back clicks on the same pixel position, and another 22 participants who failed the engagement trials, leaving us with a final sample size of 130. These participants were between 20 and 71 years of age ($\mu = 37.3, \sigma = 10.5$). 108 (83.1\%) participants used mouse clicks and 21 (16.1\%) participants used a trackpad. One participant used a touchscreen. Participants were compensated \$1.75 for their time. On average, the experiment took 7.8 minutes.

Crowdsourcing platforms exchange some control for ecological validity: sizes and colors may be affected by the display and environment used by each participant. However, this variety reflects visualization viewing in practice and has been shown to produce reliable models in past visualization research \cite{heer2010crowdsourcing,kosara2010mechanical,Kim2018c,szafir2017modeling,Smart2019b}. 

%% file: results.tex
\label{sec:results}

\ac{[R3] We introduce our notation here to avoid confusion.}
\ds{We report inferential statistics, means, and 95\% bootstrapped confidence intervals (means $\pm$ 95\% confidence intervals) for relevant effects in accordance with guidelines for transparent statistical communication \cite{dragicevic2016fair}. Means and confidence intervals are reported in pixel values: our stimuli size was 500\textit{px} by 500\textit{px}.}

\subsection{Error}
Figure \ref{fig:errors} summarizes error rates across the experimental conditions. We did not find any significant differences of error in the lightness experiment.  
In the size experiment, there was a significant interaction effect of correlations and size range on error rates ($F(4, 63) = 3.02, p < .02$). 
Tukey's HSD reveals that error rates \ds{rose dramatically} in the high-correlations, high-size range condition ($\mu=70.2px\pm5.0$).

We used Dunnett's Method to compare errors in the control condition and the three experimental conditions for both experiments. We found no significant differences between the control condition and the three lightness conditions. However, each size condition \ds{introduced significantly greater error rates than} the control condition (narrow size ranges: $p < .05$; middle size ranges: $p < .005$; large size ranges $p < .0001$).

\subsection{Bias}
\label{pull}
\label{sec:bias}

\todo{Deleted a couple sentences}
While the previous section showed rates of \emph{precision} that were comparable across conditions, \emph{bias} in participants' responses may still increase, making error rates a less reliable measure in evaluating position mean perception. 

To illustrate, assume that participant responses in the control condition were normally and randomly distributed around the true mean, and that size or lightness pulled these responses northeast.
In this case, some responses (i.e., those already northeast of the true mean) are pulled further away from the true mean, potentially increasing the error rates.
However, some responses (i.e., those southwest of the true mean) are simultaneously pulled closer to the true mean; these displacements can effectively balance out the above shifts in error rates. 
  
\begin{figure*}[!th]
	\centering
	\includegraphics[width=\textwidth]{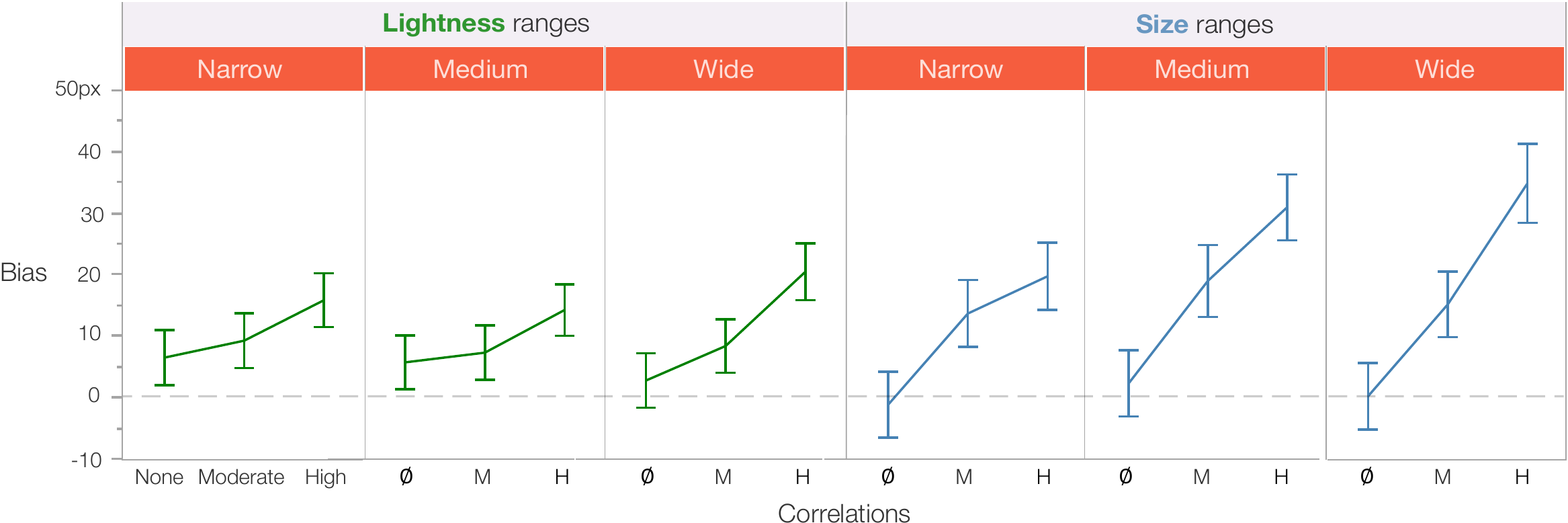}
	\caption{Mean bias across encoding type, encoding range, and correlations. Note the y-axis represents pixel values, and our stimuli size was 500\textit{px} by 500\textit{px}. Errors in both lightness (green) and size (blue) exhibited systematic biases towards locations of larger or darker points. These effects were amplified by increased correlations. Error bars represent 95\% CIs. }
	\label{fig:pulls}
\end{figure*}

\ds{Correlations between the distractor encoding and position created a visual gradient (from light-to-dark, or small-to-large) along one of the four diagonals.} We measured bias as a function of the amount of \ds{signed} error displaced along each scatterplot's direction of gradient. Overall, position means were biased toward the direction of increasing size or darkness, with the magnitude of the bias increasing with correlation between the distractor encoding and position. Figure \ref{fig:pulls} summarizes the results.
\todo{Noting the size of our stimuli [R4]}

\paragraph{Lightness: } 
While we did not find a significant effect of increasing lightness range alone ($F(2, 67) = 0.77, p < .5$), bias in the perceived mean increased as the correlation between position and lightness increased ($F(2, 67) = 4.51, p < .02$). 
Highly correlated scatterplots with a wide lightness range ($\mu=22.2px\pm5.4$) were significantly more biased than all other conditions except the other highly correlated scatterplots (narrow range: $\mu=17.1px\pm5.4$; middle range: $\mu=15.8px\pm5.4$). 
The highly correlated conditions with narrow and middle ranges displayed significantly higher bias than uncorrelated data 
with moderate ($\mu=7.3px\pm5.4$) and wide ($\mu=17.1px\pm5.5$) lightness ranges.

\paragraph{Size: } 
\ds{While we did not find a significant effect of increasing size range alone} ($F(2, 63) = 0.47, p < .7$), 
there was a significant interaction effect between increasing correlations and increasing size levels ($F(4, 63) = 2.44, p < .05$). 
Increasing correlation between size and position led to significant increases in bias ($F(2, 63) = 15.22, p < .0001$). 
Specifically, in the high correlations conditions, middle ($\mu=32.91px\pm9.6$) and wide size ranges  ($\mu=36.0px\pm9.7$) were significantly more biased than the narrow size range ($\mu=20.5px\pm9.7$).
Within each size range condition, if correlations were greater, bias was greater. The exception to this pattern was the 
narrow size range, where increasing correlations from low ($\mu=15.2px\pm9.7$) to high ($\mu=20.5px\pm9.7$) did not significantly affect bias.


%% file: attention.tex
Mean position estimates were biased toward locations of larger and (to a lesser extent) darker points, creating a \emph{weighted average illusion} giving more weight to areas with larger or darker marks. 
However, the above results do not give an account of \emph{why} the bias emerges. What elements of a scatterplot's design might be biasing position mean perception? 

Here, we seek to build a more predictive model of this bias as a function of visual elements and data distributions. With such a model, practitioners can better reason about design trade-offs by predicting magnitudes of bias across designs and distributions. This motivates our models of observed bias using the centroid method \cite{Sun2016c}. This technique uses linear regression to measure how much of the observed errors in participants' position mean responses is attributable to varying attention given to mark features, such as its specific size or lightness. 

\ds{Our primary goal with this analysis was to model bias in the no-correlation conditions, where size and lightness were randomly distributed. In such cases, we could not measure bias as signed errors along directions of the correlation gradient as there was no correlation gradient to project against.} Models of this condition can provide a baseline rate of bias for any given trivariate scatterplot. Since it is less likely that random scatterplots contain global features like a texture density gradient, \ds{which is a potentially confounding factor \cite{Haroz2012}, this model will help us answer whether people attend to data points with certain features} differently toward the mean.

\subsection{Feature-Based Attention}
\ds{Feature-based attention is an attention mechanism that operates in a parallel, distributed manner across space \cite{Alvarez2011c}.
It does so by modulating the gain of regions in the visual cortex selective for a feature (e.g., a given color or shape) \cite{Serences2009}.
When performing a visual summary task, such as comparing average values in multiclass scatterplots \cite{Gleicher2013d}, feature-based attention helps viewers attend to each set of point marks separately before comparing their average values.}

\ds{Feature-based attention is usually considered a selective attention mechanism in visualization \cite{Szafir2016d}.
However, the distribution of attention given to objects that vary across a continuous channel, such as size or lightness, need not be selective.
Attention can be automatically attuned to one type of objects relatively more than others (e.g., large marks over small marks), while still being distributed across all objects to execute a summary task like position averaging \cite{Sun2016c}.
In other words, feature-based attention may automatically ``weigh" certain kinds of marks relatively more than others when summarizing data.}

\ds{The centroid method is a linear regression model of feature-based attention that models such weights distributed across a set of point marks. 
Much like eye tracking, the centroid computation provides a behavioral marker that acts as a proxy for studying attention. However, unlike eye-tracking, the centroid method allows us to model attention being distributed across multiple locations in parallel by estimating the weight given to certain kinds of marks based on their visual features (e.g., size or lightness levels).}

\begin{figure*}[!h]
	\includegraphics[width=\textwidth]{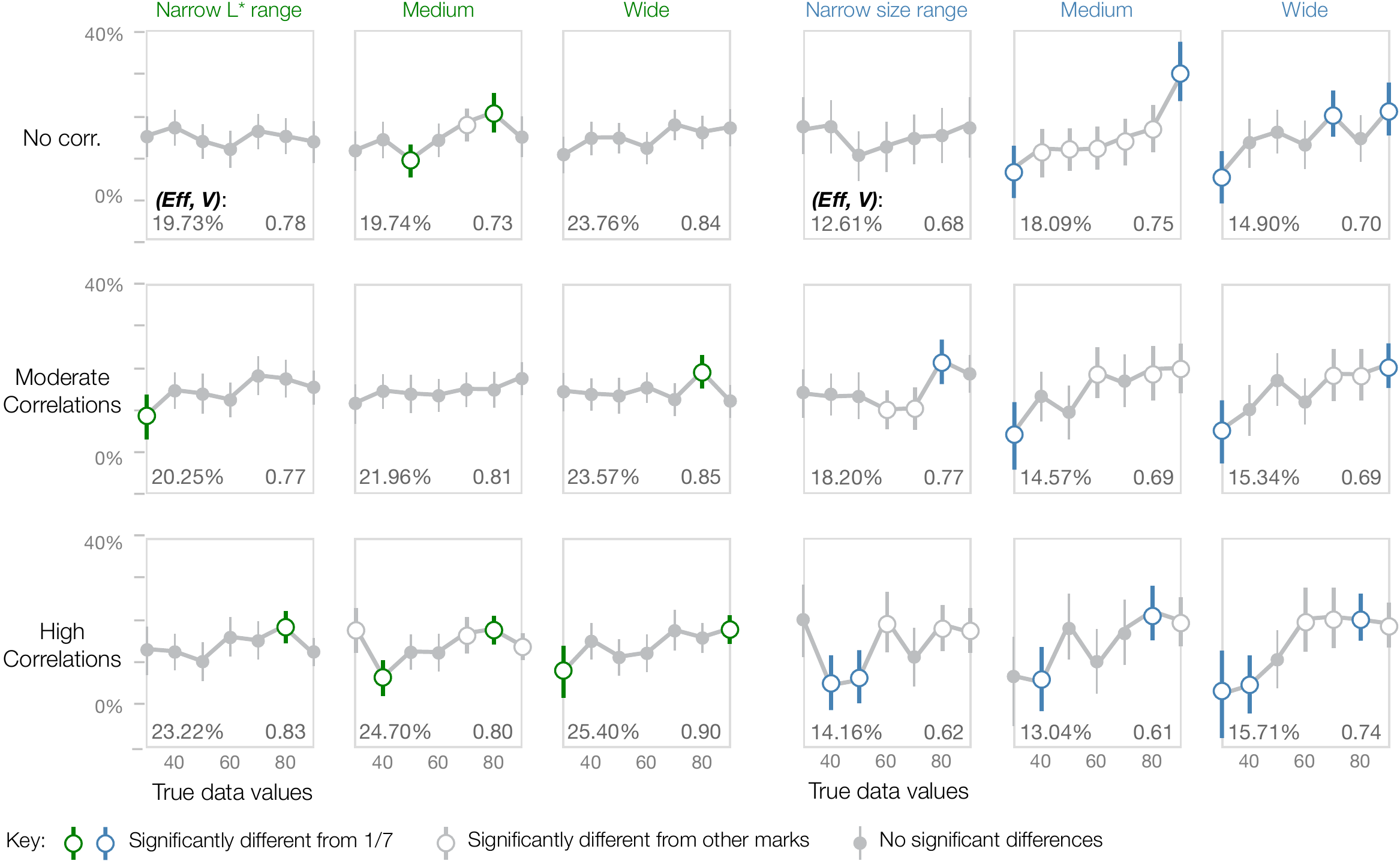}
	\caption{ Weights describing the contribution of classes of marks to the observed bias derived using the centroid method (means and 95\% confidence intervals). White marks indicate points that have significantly different weights and highlighted marks indicate where weights differed significantly from equal weighting. An upward slope indicates that people give greater weight to larger or darker marks. 
	}
	\label{fig:filters}
\end{figure*}

\subsection{The Centroid Method}
\label{sec:centroid}

The centroid method \cite{Sun2016c} defines a weight function $w(\tau)$, where $\sum^{\tau}w(\tau) = 1$, as the weight given to each item of type $\tau$.
This weight function $w$ is called the \emph{attention filter}, a model of feature-based attention \cite{Carrasco2011}---our ability to attend to items with specific visual features more, or less, than other items---which may cause the bias towards larger and darker points. This attention filter corresponds to the visual `weight'\todo{Typo [R2]} 
of individual marks. We discuss the role of \ds{feature-based} attention in interpreting our results in \S \ref{sec:fba}.

We model a participant's mean estimate $(R_{t,x}, R_{t,y})$ on scatterplot trial $t$ as
\begin{center}
	$ R_{t,x} = V\mu_{t,x} + (1-V)x_{default} + Q_{t,x} $
	$ R_{t,y} = V\mu_{t,y} + (1-V)y_{default} + Q_{t,y} $
\end{center}
where $\mu_{t,x}$ and $\mu_{t,x}$ are coordinates of the true mean position of the points; $V$, where $0 \leq V \leq 1$, is the \emph{Data-Drivenness} parameter---a measure of how much participants depended on a default location $(x_{default}, y_{default})$ (e.g.,\todo{Style [R2]} 
the center of the graph) rather than true point positions to compute mean position---and $ Q_{t,x} $ and $ Q_{t,y} $ are independent and normally distributed random response errors.

We can define $\mu_{t,x}$ and $\mu_{t,x}$ as the weighted sums of mark coordinates divided by the sum of all weights:
\begin{center}
	$ \mu_{t,x} = \dfrac{\sum\limits_{i=1}^{N_{stims}} w(\tau_{t,i})x_{t,i}}{\sum\limits_{i=1}^{N_{stims}} w(\tau_{t,i})} $, 
	$ \mu_{t,y} = \dfrac{\sum\limits_{i=1}^{N_{stims}} w(\tau_{t,i})y_{t,i}}{\sum\limits_{i=1}^{N_{stims}} w(\tau_{t,i})} $
\end{center}
where $\tau_{t,i}$ is the item type of mark $i$ in trial $t$, $w$ is the attention filter (the visual 'weight' of each mark), and $x_{t,i}$ and $y_{t,i}$ are coordinates of mark $i$ in trial $t$. 

The process for generating point and interval estimates for $w$ is outlined in Appendix 1 and 2 of Sun et al. \cite{Sun2016c}. The point estimation finds the maximum likelihood estimate of $w$ via linear regression. The interval estimation uses Fieller's theorem \cite{fieller1954some} for calculating a 95\% confidence interval for the ratio of two means. 
The interval estimates give the expected variance of visual weights given to each mark across the population. 

\subsection{Results}
Our stimuli contained marks of seven different sizes or lightnesses, which we used to represent our mark categories $\tau$. If participants 
weighed all marks equally regardless of size or lightness, the weight for any mark would be $1/7$ ($14.29\%$). 
This threshold provides a baseline for determining whether participants 
weighed certain marks more heavily than others. 
Figure \ref{fig:filters} summarizes the weight distributions
across conditions. 
The leftmost values within each plot correspond to the smallest or brightest marks in each scatterplot, and the rightmost values correspond to the largest or darkest marks in each scatterplot.

We found that even when \ds{size had no correlation with position (Figure 7, top row, blue)}, people weighed the smallest marks significantly less than and the largest marks significantly more 
for moderate and wide size ranges. 
When data was uncorrelated and the encoding ranges were narrow (top row, 1st and 4th columns), participants weighed marks roughly equally. 

\ds{We found that these weights varied significantly along with variations in range widths and correlations. In general, people weighed marks more heavily as they became darker or larger, correlated with the corresponding increase in bias toward locations of those marks from \S \ref{sec:bias}\todo{Typo[R4]}. 
We also found evidence that these weights, rather than a default response like clicking in the center of the graph, explain the errors we found in our data. 
The Data-Drivenness ($V$) of click responses were 81.09\% for lightness, and 69.46\% for size, comparable to those found by Sun et. al. \cite{Sun2016c} in in-person laboratory studies.}

We can use these weights to predict where people are likely to see the mean on\todo{Grammar [R3, R4]} a scatterplot (Fig. \ref{fig:teaser}). While we focused on lightness and size, this weighting approach has been used with other visual features like orientation \cite{inverso2016evidence} and hue \cite{sun2016human}. Future work could similarly extend the approach to other kinds of visualizations, such as bar charts or line charts, where position is redundantly coded using height or orientation. 

%% file: discussion.tex
We explored the relationship between size, lightness, and positional means in trivariate scatterplots. Our results show that the perceived mean of a scatterplot is biased towards larger or darker points (\textbf{\emph{H1}}). We have labeled this bias the weighted average illusion because the bias can be explained by asymmetries in weights we assign to marks based on irrelevant properties like size and color. We found that these effects were robust to training and increased as the structure in the data and range of size or lightness increased (\S\ref{sec:results}). 
Bias always increased as correlations between position and the third data dimension increased. This bias was directed toward areas of larger or darker points and, in the strongest conditions, caused people to misread the average by 35 pixels, supporting \textbf{\emph{H2}}. Widening size ranges also affected bias as correlations increased, partially supporting \textbf{\emph{H3}}. While these effects were stronger for size than lightness, they demonstrate the predictability of this bias as a function of data patterns and design choices (Figure \ref{fig:pulls}). 

\subsection{Design Implications}
Our results indicate that, despite classical guidelines that position is robust to other encodings \cite{ware2019information}, adding additional data to scatterplots can interfere with and even bias people's abilities to reason over position.  
These biases likely extend to other visualizations using both size and position to communicate data, such as Augmented Stripplots\cite{rensink2014prospects}. While we explore this phenomena in the context of mean judgments, bias may occur in other summary tasks that rely on similar visual processes, like variance estimation and correlation. 
Designers can use the observed bias and weights to predict when this bias might occur and use alternative design strategies, like providing reference lines or explicit summary values, to support critical summary tasks. 

Size is a more precise channel for communicating data than lightness \cite{Cleveland1984}, but 
our study indicates that size interferes with position summaries more significantly. 
Error rates in the lightness experiment were often comparable to those in the control condition.
Further, bias in bubble charts may be much greater in practice than demonstrated in our results where our tested ranges were limited to prevent occlusion. Designers should consider these trade-offs given their target audience, datasets, and the visualization tasks at hand: if assessing global properties, such as means or variances\todo{Typo [R1]}, are more important than comparing values, lightness may provide a more robust channel. If interpreting individual values is more critical, size may be preferable.


\ac{Discussion of how designers can utilize the centroid method}
\ds{
	Using the centroid paradigm, designers can run exploratory self-studies to quantify the degree to which readers can summarize relevant data when distracting categorical (e.g., hue \ds{palettes} \cite{sun2016human, Munzner2018c}\todo{Clarifying what we meant by hue, and adding a citation to justify that hue is categorical [R1]}) or continuous channels (e.g., size or lightness) are present. 
	When the relevant and distractor channels may not seem separable, weights modeled using the centroid method can be used to estimate a baseline rate of bias. 
	For example, 
	by simply computing the average of mark locations weighed according to our attention filters presented in Figure 7, designers can predict which data points will be seen as being above or below average.
	This is how the predicted perceived means in Figure \ref{fig:teaser} (open circles) were computed. 
	Developing a more extensive set of models covering a larger range of designs could provide significant predictive power for predicting bias in information design.
}

Both current and prior work suggest that readers can be trained to mitigate this bias in small collections of objects. While our results exhibited a consistent bias, there can be significant individual differences in how much weight people give to each mark during averaging that can be reduced with sufficient training \cite{Drew2010c}. 
In pilot studies, we found that without training, people exhibited a wider range of bias magnitudes but these biases converged over time. We account for this in our study with our training phase, where participants receive feedback for 18 trials before entering the formal study. However, these findings suggest the possibility that certain kinds of biases can be mitigated by training. 


\begin{figure}[h!]
	\centering
	\includegraphics[width=\columnwidth]{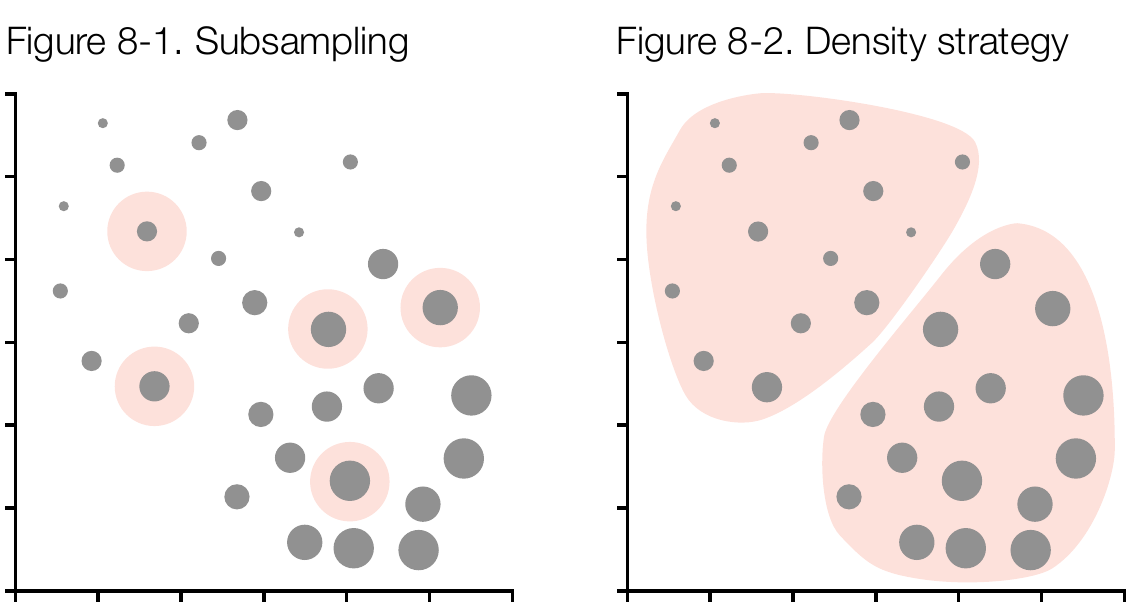}
	\caption{Potential strategies employed by participants. (Left) Participants could have been focusing on a small subset of point marks, shortcutting the need to attend to all points. This would bias the mean position if darker or larger marks are more likely to be focused on. (Right) Participants could have segmented highly correlated scatterplots into sets of smaller points and sets of larger points, then computing the midpoint of those segments weighted by estimates of density in each segment. This strategy amounts to asking the question "How many points are here, as opposed to there?" instead of directly computing the mean position. Although the segment with larger marks might seem to contain more data, both segments contain 15 marks.}
	\label{fig:strategies}
\end{figure}  

\subsection{Potential Basis for the Weighted Average Illusion}
\label{sec:strategies}
By modeling how different elements of a design may bias mean estimates, we can generate new hypotheses for visualization psychology that allow us to anticipate \emph{why} this bias might arise and \emph{when} it may impact other visualization types or tasks. 
Many ensemble coding strategies, proxies, and heuristics can support or combine to help people compute visual summary statistics. 
We expand on what the centroid method results (\S \ref{sec:centroid}) tell designers about attention in trivariate scatterplots and propose two possible mental shortcuts that people could be using to compute mean positions: 1. sampling only a subset of marks that could be held in working memory (Figure 8-1), and 2. comparing texture densities between spatial segments of the data (Figure 8-2).

\subsubsection{Local Basis: Feature-Based Attention}
\label{sec:fba}
\ds{The position mean is a reliable behavioral marker for studying feature-based attention in vision science \cite{Sun2016c} and information visualization \cite{Gleicher2013d}.}
Feature-based attention implies that attention is \ds{unevenly distributed across space, skewing towards marks with certain visual features \cite{Carrasco2011}.
The attention filters modeled with the centroid method quantify this skew \cite{Sun2016c}.}
When reading visualizations, we rely on feature-based attention to, for example, make judgments about different data categories \cite{Gleicher2013d} or search for relevant data \cite{haroz2012capacity}. 


\ds{Although we did not use eye tracking in our study (as discussed in \S\ref{sec:limitations}), prior work already confirmed that dark and large marks are more salient than light and small marks \cite{Healey2012}.
This implies that participants' eye movements may have reflexively saccaded to the locations of dark or large objects as soon as each scatterplot was presented.
Such behavior might have a priming effect on computing the mean position of scatterplot marks, causing feature-based attention to be distributed more towards larger or darker marks and potentially leading to the skewed attention filters modeled in Figure 7.}

\ac{Better transition to next two subsections [R1]}
However, if feature-based attention was the only factor at play, we would see the same attention filters in each correlation condition, varying only as a function of increasing size or lightness ranges. 
Instead, we found that the attention filters varied with increasing correlations in the data as well, \ds{indicating there is more to the observed bias than simple pop-out effects. \todo{Revise sentence}
People may alternatively be using subsampling or density-driven strategies (Figure 8). We discuss these strategies below as both potential explanations for the observed bias as well as opportunities to inform future designs. }

\todo{Remove paragraph}

\subsubsection{Hyperlocal Basis: The Subsampling Strategy}
\label{sec:subsampling}

Myczek and Simons \cite{Myczek2008} demonstrated 
that  subsampling four objects from a larger collection\todo{Typo [R4]} 
can 
account for performance on size averaging. However, other studies show that people may need significantly more information to make sense of eight or more objects
\cite{Alvarez2008d}, far fewer than the number of marks in a typical visualization, with later work estimating that the number of samples must be closer to the square root of the number of marks \cite{whitney2018ensemble}.  
Using the centroid method, we can compute the minimum portion of marks, called the \emph{Efficiency}, or \emph{Eff}, required to converge onto our models in \S\ref{sec:centroid} ($w$, $V$, $x_{default}$ and $y_{default}$).  In other words, this measure of efficiency quantifies the minimum number of objects that would have to be used in computing position mean to achieve the same level of performance achieved by the participants.

Given the residual sum of squares $SS_{Residual}$ derived using the centroid method, the standard error $\widehat{\sigma}$ provides an unbiased estimate of the standard deviations of the random response errors $Q_{t,x}$ and $Q_{t,x}$:

\begin{center}
	$\widehat{\sigma}=\sqrt{\frac{S S_{\text {Residual }}}{d f}}$
\end{center}

We assume as in prior work \cite{Alvarez2008d} that all error in $\widehat{\sigma}$ was due to the number of items unattended to by participants. To compute \emph{Eff}, we use the technique from Sun et al. \cite{Sun2016c} to calculate the variance of the difference between participants' responses and the predicted responses using $w$, $V$, $x_{default}$, and $y_{default}$ after removing $N$ marks. Starting by deleting only $N=1$ mark from each stimulus scatterplot, this variance is iteratively computed for each possible number of marks that can be deleted until a single mark is remaining ($0 < N < 30$), terminating once the variance is greater than $\widehat{\sigma}$.

Averaging over the experimental conditions, our models in Figure \ref{fig:filters} could have been achieved by attending to as little as 22.48\% of marks (around 7 marks) in the lightness experiment, and 15.18\% of marks (around 5 marks) in the size experiment, aligning with the square root of the number of marks predicted in Whitney \& Leib \cite{whitney2018ensemble}.
\ds{These results may imply that people attend to a small number of marks 
	when summarizing trivariate scatterplots.}
Our measured bias could arise if dark or large marks capture more attention and are disproportionately represented in participants' subsamples. \todo{Add sentence}
If people are subsampling marks, then\todo{Typo} 
we anticipate that this bias may arise in other visualizations that use discrete marks, such as bar charts.
This bias may be reduced by using continuous representations like kernel-density estimation \cite{eilers2004enhancing}.\todo{Style [R4]} 

\subsubsection{Global Basis: The Density Strategy}
\label{sec:density}

Attention filters in \S\ref{sec:centroid} varied as a function of both encoding ranges and correlations, suggesting that clustering darker and larger marks together may introduce other factors that increase the observed bias. 
One factor may be that the density across different spatial segments also pulls the perceived mean.

We use the term \emph{density} to refer to any of three mechanisms possibly involved numerosity estimation in a group of marks: subitization (when discriminating 10 or fewer data points), estimation (for more than 10 data points), and low spatial frequency features such as texture density or contrast energy \cite{Pome2019c} (for review, see Picon et al. \cite{Picon2019c}). People can estimate the densities of overlapping dot textures of different colors in parallel \cite{Halberda2006} and visualization techniques leverage density to deal with challenges like overdraw \cite{eilers2004enhancing,mayorga2013splatterplots}.

When a scatterplot depicts a third data dimension 
and this dimension is correlated with position, 
the resulting clusters will share the same visual features. For example, a bubblechart will have clusters of mostly large and mostly small dots. 
When clusters emerge, the visual system may immediately segment the scene and then estimate the density of the resulting segments \cite{Franconeri2009c}.

\ac{[R1] Removing reference to Melcher \& Kowler as it isn't relevant} 
Some participants might have
computed the midpoint of those centroid estimates, weighted by estimates of density in each segment.
Prior work confirmed that large or dark points clustering together within an area creates illusions of higher point density \cite{Hurewitz2006c,Tibber2012c,Dakin2011c,Durgin1995,Gebuis2012,Lei2018c,Picon2019c,Morgan2014b}.
If people use the positions and densities of spatial segments to estimate the overall mean, these illusions will bias the mean towards clusters that contain large or dark marks. 

%% file: limitations.tex
We investigated two common visual channels---size and color---that are frequently used to encode additional information in scatterplots. 
However, the simplicity of scatterplots affords a large space of potential designs that may offer different perceptual trade-offs \cite{Sarikaya}. 
Future work might consider these alternatives to understand the robustness of mean position perception across scatterplot designs and how this bias may influence a broader range of ensemble tasks like correlation and variance estimation.
For example, the centroid method employed here could be used to understand biases introduced by a range of visual variables for both categorical and numeric data. 
Further, we measured weights along a small set of size and lightness ranges. While these weights allow us to reasonably estimate the probable location of a perceived mean over these ranges, we do not have sufficient data for a fully predictive model. Modeling weights across a wider range of factors may enable robust bias predictions across a range of encoding lengths and designs. 

Our experimental stimuli were generated according to Poisson disk sampling, creating random dot textures.
While these textures contain a range of clusters and structures, real datasets often have additional global features that can cause data points to more strongly cluster. 
These factors could introduce additional biases in realistic scatterplots and may limit the generalizability of the computed weights. 
Additionally, our data generation approach led to scatterplots where the direction of correlation was slightly displaced from the true diagonal. Although these displacements are small and randomly distributed, since we define bias as the magnitude of the error vectors projected onto the diagonals, the bias illustrated in Figure \ref{fig:pulls} may underestimate the true bias. 

While our crowdsourcing study led to consistent results, the attention filters derived in \S\ref{pull} may vary across individuals. 
For example, certain people can discount the contribution of peripheral objects toward the mean \cite{Drew2010c}. These individual differences may shift the ways people reason about means, especially for visualizations targeting data within specific areas of expertise \cite{kastens2016geoscience}. Disciplines may exaggerate bias by introducing semantic factors such as context or risk that add mental ``weight'' to critical data. Future work might leverage the centroid method to model these individual differences under varying data contexts to the influence of disciplinary knowledge and other factors on bias. 

\ds{Lastly, we identified two potential strategies that might explain the observed bias. Given that feature-based attention operates prior to voluntary eye movements \cite{Mazer2011} and constraints of the COVID-19 pandemic, we did not incorporate eye tracking in our study. However, eye-tracking may provide further insight into these strategies. 
Future work should combine the centroid computation and eye tracking, both state-of-the-art behavioral markers for studying visuospatial attention, to investigate the subsampling and density hypotheses.
}

%% file: conclusion.tex
Our abilities to estimate summary statistics from scatterplots may be sensitive to attributes of the scatterplot design and underlying data \cite{Kim2018c}. We conducted a crowdsourced experiment investigating how irrelevant channels, the ranges of those channels, and their relationhips with the position channel may shift the position mean. We found that 
the perceived mean position in a scatterplot is biased in the direction of larger, and to a lesser extent, darker marks. We model the contribution of different elements of a visualization design towards this bias using the centroid method, a statistical technique from vision science, offering designers a way to predict weights people give to a given mark. 

Our results raise new opportunities at the intersection of visualization and vision science: we elucidate a systematic bias that
gives insight 
into feature-based attention in visualization as well as a variety of ensemble strategies that can be employed in visualization interpretation and decision making. 
For designers, modeling bias as a function of design choices and data patterns allows designers to avoid misleading practices and 
experienced graph readers to notice and correct for potential bias. 

\ds{
Research in visualization biases tend to develop in parallel with research in visualization literacy, since biases often arise when readers are not familiar with a visualization \cite{Mansoor2018}.
The weighted average illusion adds to the growing literature of biases \cite{dimara2018task, Xiong2020c, WittWarden2020, Procopio2021, Dimara2019, CaleroValdez2018, Alexander2018} that may affect interpretations of even the most familiar visualizations.

In our pilot studies, even visualization experts had a hard time avoiding this bias, likely naming justifications for weighing point marks differently, such as: ``What if the size channel represented population size?"
We emphasize that this illusion is a misjudgment regardless of what lightness or size represents, since weighted scatterplot means change with the ranges of visual channels a designer chose to map the data onto.
By elucidating these biases, we can both promote honest communication in visualization practice, and guide the development of data literacy for everyone.
}